\documentclass{an}

\usepackage{epsfig}
\usepackage{graphics}

\newcommand{\be}{\begin{equation}}
\newcommand{\ee}{\end{equation}}

\begin{document}

\title{Mattig's relation and dynamical distance indicators
}
\author{P. Teerikorpi\inst{1} \and Yu.V. Baryshev\inst{2}}
\institute{Tuorla Observatory, Department of Physics and Astronomy,
 University of Turku, 21500 Piikki\"{o}, Finland \and
 Astronomy  Department, Saint Petersburg State University, Staryj Peterhoff, 198504, St Petersburg, Russia}
%\date{Received / Accepted}

\titlerunning{Mattig's relation and dynamical distance indicators}

\authorrunning{P. Teerikorpi and Yu.V. Baryshev}

\received{}
\accepted{}
\publonline{later}


\abstract{
We 
discuss how the redshift (Mattig) method in Friedmann cosmology relates to
dynamical distance indicators based on Newton's gravity (Teerikorpi 2011).     
It belongs to the class of indicators where the relevant length inside the system is the distance itself (in this case the proper metric distance). As the Friedmann model has
Newtonian analogy, its use to infer distances has instructive similarities to classical dynamical distance indicators.
In view of the theoretical exact linear distance-velocity law, we emphasize that it is conceptually
correct to derive the cosmological distance via the  route: redshift (primarily
observed) $\rightarrow$ space expansion velocity (not directly observed)  $\rightarrow$ metric distance (physical length in "cm").
Important properties of the proper metric distance are summarized.
}

\maketitle

\section{Introduction}

Newtonian acceleration $\ddot{r}$ of a particle in the force field of a spherical
mass $M$
\begin{equation}
\frac{1}{r^2} = - \frac{1}{G} \times \frac{\ddot{r}}{M}\,.
\label{equation1}
\end{equation}
offers known ways to infer the distance $d$ of a gravitating system. 
Such dynamical distance indicators often belong to "physical" (Sandage et al. 2006) or
"one-step" (Jackson 2007) methods, in contrast to relative distance indicators
such as standard candles and rods. The latter give luminosity and
angular size distances. The distance type from physical indicators depends on the method; essential is that the distance is directly obtained in cm, with no midway distance as a reference.

Note that
 $r$ in Eq. (1) can be either a length {\it within} a remote system whose
distance is wanted or it can be the distance itself from the observer to the system.   

\subsection{When $r$ is within a distant system}
 
When the acceleration can be related to an observable velocity quantity $V$,
Teerikorpi (2011; T2011) grouped dynamical distance indicators
according to how the $M/d$ degeneracy in the equation
$M = c_1 \theta d V^2\, $
is overcome if one can express $M/d$ in another form containing  known quantities
and the unknown $d$ ($\theta$ is the angle the size $r$ is observed at).
T2011 discussed cases $M \propto d^n$ where
$n=0, 2$ or 3, and the distance $d$ can be determined. 

\subsection{When $r$ is the desired distance $d$ from us}

In some cases the length in Eq.(1) is the distance $d$ 
from us to an object, as in a binary system with the observer on one of the bodies.
Idealized examples could be the Earth-Moon and Sun-Earth pairs.
Assuming a circular orbit, the radial acceleration
 leads to $d = \frac{GM}{V^2}$, with the orbital speed 
from the period $T$ as $V = \frac{2\pi d}{T}$. This leaves
$M \propto d^3$ so one needs an auxiliary relation $M \propto d^n$, where $n\neq 3$.
For the Earth-Moon pair one writes
$G M_{\rm E}= gr_{\rm E}^2$ (so $n = 0$). For the Sun the gravitational redshift $z_{\rm g}$  gives
(T2011)
 $GM_{\rm sun} = z_{\rm g}c^2 \theta_{\rm sun}d$ ($n = 1$). 

Such cases involved circular orbits. In the Galaxy--M31 pair and in the flow of dwarf galaxies away
from the Local Group the motion is (nearly) radial and one cannot directly infer the acceleration in Eq.(1). However, with data on the mass and "bang time" one can derive, by Tolman-Bondi-type calculations, the expected
distance-velocity relation that gives the distance if the radial velocity is measured (e.g., Gromov et al. 2001).

A supreme distance indicator of this category with radial motion is the Friedmann universe itself. Its special features are discussed below.

\section{The Friedmann universe}

The spheres cut from this world model are known  to expand in  a Newtonian way
(McCrea \& Milne 1934). In terms of internal hyperspherical coordinates, with the scale factor $a$ chosen to be dimensionless and the comoving radial coordinate $r_{\rm c}$ so that
the metric distance is $d = a r_{\rm c} = r_{\rm c}$ at the present time ($a =1$):
 \begin{equation}
\frac{1}{(ar_{\rm c})^2} = - \frac{1}{G} \times \frac{\ddot{a} r_{\rm c}}{M_{\rm eff}}\,,
\label{equation3}
\end{equation}
where the effective mass $M_{\rm eff} = \frac{4\pi}{3}(ar_{\rm c})^3(\rho_{\rm m}- 2\rho_{\Lambda} )$
 ($\rho_{\rm m}$ is the density of the  gravitating matter and $\rho_{\Lambda}$ is that of the $\Lambda$ term).
We note that Eq.(2) is exact in the Friedmann model, not an approximation valid, say, 
at short distances (Baryshev 2008; Baryshev \& Teerikorpi 2012). 

Here the dynamical equation is, of course, Einstein's field equation.
With the symmetries underlying the world model it leads to Friedmann's equations (and Eq.(2)).
 The puzzling similarity of Eq.(2) to Eq.(1) has led to discussions on its meaning (e.g., Layzer 1954)
and if it could throw light on conceptual problems in Friedmann cosmology (e.g., Baryshev 2008). 
Here we simply ask if the analogy extends to how one obtains the (well-known) distance indicator in the Friedmann world.

We cannot measure the acceleration of space expansion $\ddot{a}r_c$
between us and the remote object. 
However, we can replace $\ddot{a}$ in Eq.(2) by $-q \dot{a}^2 /a$, where $q$ is the deceleration parameter, constant on all scales
$r_{\rm c}$ at the epoch $t$ (fixed $a(t)$), due to the system's regularity. So
 \begin{equation}
(\dot{a}r_{\rm c})^2 =  \frac{G}{q} \times \frac{M_{\rm eff}}{ar_{\rm c}}\,,
\label{equation3}
\end{equation}
which resembles the epressions often appearing for classical dynamical distance indicators (T2011). Writing
$V =\dot{a}r_{\rm c}$ and $d=ar_{\rm c}$, we see also here the $M/d$ degeneracy. It can be overcome, as
$M_{\rm eff} \propto d^3 $, so $n=3$. Inserting the expression for $M_{\rm eff}$ into Eq.(3) one gets
 \begin{equation}
V=\dot{a}r_{\rm c} =  (\frac{\gamma}{q})^{1/2} \times {ar_{\rm c}}=H(t) d\,,
\label{equation4}
\end{equation}
the linear distance-velocity law in the Friedmann world. Here $\gamma=
 \frac{4\pi G}{3} (2\rho_{\Lambda} - \rho_{\rm m})$
is constant for a fixed scale factor. 
It is this law that offers a route to the distance.

In a landmark study, Harrison (1993) underlined that the relation
$V = Hd $
is exact. $V$ is the rate at which the (proper) metric distance $d$ between
two objects at rest in space grows.
The Hubble
constant $H_0$ fixes this directly unobservable law in its current state.
 
The expansion
rate $V$ cannot be directly measured. In classical indicators if the radial velocity is needed it is inferred using the Doppler effect. The expansion rate between us and the object is also got from the spectral line shift but now using the Lema\^{i}tre (1927) effect that gives the exact sense of the cosmological redshift.

\section{From redshift to velocity to distance}

In his classic paper, Mattig (1958) solved the task of deriving the formula for the luminosity distance for
a given redshift. His aim was to express the apparent magnitude
of a standard candle as a function of
redshift, and the focus was not on the metric distance and the exact linear expansion law (which had not yet been adequately discussed at the time).\footnote{Wolfgang Mattig derived his result as 
student when he had to give a talk on cosmology for his
graduate exam. A letter to the present authors reveals that during the short preparation time "I also studied Heckmann's book on 'Theorien der Kosmologie' and I found 
it extremely insufficient that the relations $z(m)$ and $N(m)$ are given in series expansions. 
I tried to find a closed form for the simplest case, zero cosmological constant and flat space. I succeeded
within the two weeks and got through the examination with a good result." He also noted:
"the reaction of the cosmological community was nearly zero, only Allan Sandage discussed my relations in ApJ 133 (1961). He rendered accessible my results [which] were published
in German, in an East-German journal." The letter is printed
in Baryshev \& Teerikorpi (2002). For interesting remarks, see Sandage (1995).}

After Harrison's (1993) work, it is natural  to underline the exact link between the expansion velocity and the metric distance. 
It is also instructive  and logical when considering  the expansion law  as a distance indicator, to derive first the present moment $V(z)$ for the redshift $z$ and then 
divide it by $H_0$. The mathematics is essentially as in
Mattig's original way. 

In general, the light emitted from a galaxy having the redshift $z$, has
gone through a comoving radial coordinate interval $r_c$ before having reached us.  
Then one can write for the present rate of increase of the metric distance between us and the galaxy,
using $a\mathrm{d}r_c = c\mathrm{d}t$:
\be
V(z) = (\frac{\mathrm{d}a r_c}{\mathrm{d}t})_0  = (\dot{a})_0 c  \int_{a_0/(1+z)}^{a_0} \frac{\mathrm{d}a}{a \dot{a}} \, .
\ee 
One may express the derivative $\dot{a}$ in terms of $a$ using the first  Friedmann equation.
Changing the variable from $a$ to $z$, noting  how 
 $\rho_{\rm m}$ and $\rho_{\Lambda}$ depend on $a$, and dividing by $H_0$  gives the known formula
\begin{eqnarray}
d(z) &=& \frac{V(z)}{H_0} \nonumber \\
       &=& \frac{c}{H_0} 
\int_{0}^{z} \frac{\mathrm{d}x}{(\Omega_{\rm m} (1+x)^3 + \Omega_{\Lambda}
+\omega (1+x)^2 )^{\frac{1}{2}}}
\end{eqnarray}
where $\omega = 1-\Omega_{\rm m} - \Omega_{\Lambda}$ ($= 0$ for flat space).
Only for the flat pure $\Lambda$ model the expansion velocity is  the "classical Doppler" $V(z) = cz$.
Eq.(6) can  be generalised to
include energy transfer between the matter components (Teerikorpi et al 2003; Gromov et al 2004).

\section{Concluding remarks}

We showed the steps from redshift to metric distance in order to sum up
the idiosyncracies of the redshift method 
 where the system is the universe as described by the Friedmann model. 
This machinery as a dynamical distance indicator involves tightly packed
all the ingredients of cosmological physics, such as the Cosmological Principle, general relativity, space expansion
(instead of motion in static space), and the  Lema\^{i}tre redshift effect.
At the same time, it is instructive to see that each step has analogies with classical dynamical distance indicators (Eqs.(2)-(5)).

Remarkably, when the density parameters of the Friedmann model  $\Omega_{\rm m}$ and $\Omega_{\Lambda}$ are fixed, a single
observation of the dimensionless spectral redshift allows one to derive the current space expansion rate (in cm/s)
between us and the object in question. Then, the exact distance-velocity relation (with
the Hubble constant $H_0$) leads
to the  proper metric distance, as expressed in units of cm. This is the correct, 
straightforward and simple way to speak about the derivation  of the distance within the
standard cosmology. 

We find it useful to finish with 
some notes on the end result, the metric distance. 
McVittie (1974) wrote that "distance is a measure of remoteness". In his analysis of different
distance types that appear in cosmology he did not prefer any one of them above the others. However, with the Friedmann model as the basis of cosmology, there are good reasons to
view the  "tape measure" metric distance $d$ as basic. It comes close to our ordinary physical conception of distance, even though it cannot be directly measured.

This distance appears in the exact expansion law. It is also additive, unlike, say, the luminosity distance.
Also, for $z = \infty$ it gives the particle
horizon (e.g., $2 \frac{c}{H_0}$ in the Einstein-de Sitter model ($\Omega_{\Lambda}= \omega = 0$ in Eq.(6)).

 The metric distance is the
backbone distance related in known ways to the work horses of practical cosmology: luminosity and angular size distances.
The link is especially simple in flat models, where  $d_{\rm lum} =(1+z)d$ and 
$d_{\rm ang} =(1+z)^{-1}d$ and the metric distance $d$ is the geometric mean of those other distances.\footnote{In non-flat models 
$d_{\rm lum} = (1+z)d_{\rm ext}$ and $d_{\rm ang} = (1+z)^{-1}d_{\rm ext}$, where
$d_{\rm ext}$ is the external metric distance (so termed by Baryshev \& Teerikorpi 2012), which Mattig (1958)
derived for zero-$\Lambda$ dust models as his Eq.(12). It coincides with the internal distance $d$ in flat models, being otherwise related to $d$ as
$d_{\rm ext} (z) = R_{\rm H} (k\omega)^{-1/2}sn((k\omega)^{1/2}\frac{d(z)}{R_{\rm H}})$, where
$sn = sin$ when $\omega > 0$ (and $k = 1$) and $sn = sinh$ 
when $\omega < 0$ (and $k = -1$). $R_{\rm H}$ is the Hubble distance
$c/H_0$.}

 Finally,
the current metric distance is the best "measure of remoteness" characterizing the location of a distant
object in the galaxy universe. So, if its metric distance is 800 Mpc, we at once know that
it is 1000 times more distant than the Andromeda nebula.
 Therefore, the metric distance is also the choice, say, in science news about very distant
galaxies, preferably with the familiar light-year as unit.

The metric distance in light-years may be so large that its numerical value exceeds the age of the universe in years. This may disquiet an attentive reader, wondering how the light had enough time to cross such a distance. Then the balloon analogy by Eddington (1930),  illustrating the Hubble
law and the changing metric distance, also offers an intelligible solution to this "age paradox". In fact, the "light travel distance" is obtained by weighing the integrand in Eq.(6) by the redshift factor  $(1+x)^{-1}$. 

{\it \bf Acknowledgements} We are grateful to Wolfgang Mattig who kindly informed us on the history of his 1958 paper in Astronomische Nachrichten. We also thank the referee for a careful reading of the manuscript. Yu.B. thanks the Saint-Petersburg University  for financial support
to this work (grant 6.38.18.2014).

{}

\end{document}